\documentclass[10pt, twocolumn,superscriptaddress,preprintnumbers]{revtex4}%
\usepackage{epsf}
\usepackage{amsmath}
\usepackage{amssymb}
\usepackage{epsfig}
\usepackage{latexsym}
\usepackage{amsfonts}
\usepackage{dcolumn}
\usepackage{varioref}
\usepackage{ifthen}
\usepackage{graphicx}
\usepackage{amssymb,amsmath,amsthm}%
\providecommand{\U}[1]{\protect\rule{.1in}{.1in}}
\begin{document}
\preprint{UCT-TP-287/11}
\title{Hadronic contribution to the muon g-2: a theoretical determination}
\author{S. Bodenstein}
\affiliation{Centre for Theoretical \& Mathematical Physics, University of Cape Town,
Rondebosch 7700, South Africa}
\author{C. A. Dominguez}
\affiliation{Centre for Theoretical \& Mathematical Physics, University of Cape Town,
Rondebosch 7700, South Africa}
\author{K. Schilcher}
\affiliation{Centre for Theoretical \& Mathematical Physics, University of Cape Town,
Rondebosch 7700, South Africa}
\affiliation{Institut f\"{u}r Physik, Johannes Gutenberg-Universit\"{a}t Staudingerweg 7,
D-55099 Mainz, Germany}
\date{\today}

\begin{abstract}
\noindent The leading order hadronic contribution to the muon g-2, $a_{\mu
}^{HAD}$, is determined entirely from theory using an approach based on
Cauchy's theorem in the complex squared energy s-plane. This is possible after fitting the integration kernel in $a_{\mu}^{HAD}$ with a simpler function of $s$. The integral determining $a_{\mu}^{HAD}$ in the
light-quark region is then split into a low energy and a high energy part, the latter given by perturbative QCD (PQCD). The low energy
integral involving the fit function to the integration kernel is determined by derivatives of the vector correlator at the origin, plus a
contour integral around a circle calculable in PQCD. These derivatives are
calculated using hadronic models in the light-quark sector. A
similar procedure is used in the heavy-quark sector, except that now
everything is calculable in PQCD, thus becoming the first entirely theoretical calculation of this contribution.
Using the dual resonance model realization of Large $N_{c}$ QCD to compute the derivatives of the correlator leads to agreement with the experimental value of $a_\mu$. Accuracy, though, is currently limited by the model dependent calculation of derivatives of the vector correlator at the origin. Future improvements should come from more accurate chiral perturbation theory and/or lattice QCD information on these derivatives, allowing for this method to be used to determine $a_{\mu}^{HAD}$ accurately entirely from theory, independently of any hadronic model.
\end{abstract}

\pacs{}
\maketitle
\parindent 0mm \setlength{\parskip}{\baselineskip} \thispagestyle{empty}
\pagenumbering{arabic} \setcounter{page}{1} \mbox{ }
\noindent 
The value of the muon g-2 is well known as a test of the standard
model (SM) of particle physics. \cite{review}. The SM result for the anomalous magnetic moment of the muon is conveniently separated into the contributions due to QED, the hadronic sector, and the electroweak sector. A sizable theoretical uncertainty arises from the (leading order) hadronic vacuum polarization term, the second largest contribution after that of QED. A substantial effort has been made to determine this contribution from
experimental data on $e^{+}e^{-}\rightarrow\mbox{hadrons}$ and $\tau
\rightarrow\mbox{hadrons}$ \cite{Davier}-\cite{Hagiwara}. Currently, there is
some yet unresolved discrepancy between both results. Writing the muon anomaly in the SM as
\begin{equation}
a_{\mu}^{SM}=a_{\mu}^{QED}\,+\,a_{\mu}^{HAD}\,+\,a_{\mu}^{EW}\;,
\end{equation}
the leading contribution is that from QED, followed by the hadronic and the
electroweak terms. In this paper we concentrate on the leading order hadronic
contribution and discuss a new approach to its calculation entirely from
theory. The method relies on Cauchy's theorem in the complex squared energy
s-plane, after fitting the integration kernel entering $a_{\mu}^{HAD}$ with a
simple function of $s$ . In the region of the light-quark sector the method
requires knowledge of some of the derivatives of the (electromagnetic) vector
correlator at zero momentum, as well as its perturbative QCD (PQCD) behavior.
Currently, these derivatives will be obtained here from hadronic models, thus being affected by systematic uncertainties. Hence, at this stage the method cannot rival in accuracy with the standard approach of using experimental data on the vector correlator at low/intermediate energies. However, future precision determinations of these derivatives from chiral perturbation theory (CHPT) or lattice QCD would render this calculation of $a_{\mu}^{HAD}$ independent of experimental data on $e^{+}e^{-}\rightarrow\mbox{hadrons}$ and $\tau\rightarrow\mbox{hadrons}$. It must be emphasized that as a consequence of Cauchy's theorem
this method only requires knowledge  of the derivatives of the vector correlator at the origin, rather than its full expression over an extended energy region. In addition, the method allows for a
straightforward incorporation of the charm- and bottom-quark contributions to
$a_{\mu}^{HAD}$ calculable exclusively from PQCD, i.e. without the need for data on the vector correlator. This leads to the first entirely theoretical calculation of this contribution.\\
We begin with the standard expression of the (lowest order) muon anomaly \cite{review}
\begin{equation}
a_{\mu}^{HAD}=\;\frac{\alpha_{EM}^{2}}{3\,\pi^{2}}\,\int_{s_{th}}^{\infty
}\,\frac{ds}{s}\;K(s)\;R(s)\;,\label{AMUH1}%
\end{equation}
where $\alpha_{EM}$ is the electromagnetic coupling and the standard $R$
-ratio is $R(s)=3\,\sum_{f}\,Q_{f}^{2}\left[  8\,\pi\,\mbox{Im}\,\Pi
(s)\right]  $, where $Q_{f}$ are the quark charges and $\Pi(s)$ is the vector
current correlator normalized to $8\,\pi\,\mbox{Im}\,\Pi(s)=1+\alpha_{s}%
/\pi+\cdots$. The integration kernel $K(s)$ in Eq.(\ref{AMUH1}) is given by
\cite{deRK}
\begin{equation}
K(s)=\int_{0}^{1}\,dx\,\frac{x^{2}(1-x)}{x^{2}+\frac{s}{m_{\mu}^{2}}
(1-x)}\;, \label{K}
\end{equation}
where $m_{\mu}$ is the muon mass. A popular approach to compute $a_{\mu}
^{HAD}$ has been to split the integral in Eq.(\ref{AMUH1}) into a low energy
region from threshold up to $s=s_{0}\simeq(1.8\,\mbox{GeV})^{2}$, followed by
a high energy region from $s=s_{0}$ to infinity. The integral in the former
region was calculated using data on of $e^{+}e^{-}\rightarrow\mbox{hadrons}$
or (isospin rotated) data on $\tau\rightarrow\mbox{hadrons}$. The integration
in the high energy region was performed assuming PQCD. In this paper
we discuss a new approach based entirely on theoretical input. For later
convenience we split the contributions to the leading order $a_{\mu}^{HAD}$
into three pieces,
\begin{equation}
a_{\mu}^{HAD}=a_{\mu}^{HAD}|_{uds}+a_{\mu}^{HAD}|_{c}+a_{\mu}^{HAD}
|_{b} \;, 
\end{equation}
where the first term on the right hand side  corresponds to the
contribution of the three light quarks, and the second and third term refer to the charm- and bottom-quark contributions. In the light-quark sector the first step is to fit the integration kernel $K(s)$ in an interval
$s_{th}\leq s\leq s_{0}$ with a function $K_{1}(s)$
\begin{equation}
K_{1}(s)=a_{0}\,s+\,\sum_{n=1}\frac{a_{n}}{s^{n}}\;,\label{K1}
\end{equation}
with coefficients determined by minimizing the chi-squared. The upper limit
$s_{0}$ is below the charm threshold. Next, the integration range in
Eq.(\ref{AMUH1}) is split into a low energy ($s\leq s_{0}$) and a high energy
($s>s_{0}$) region where PQCD would be valid. In the former region Cauchy's
theorem is used to obtain
\begin{eqnarray}
\int_{s_{th}}^{s_0} \frac{ds}{s}  &K_{1}(s)& \, \frac{1}{\pi} \, \mbox{Im} \,\Pi_{uds}(s)  =  \mbox{Res} \left[ \Pi_{uds}(s) \frac{K_{1}(s)}{s}\right]_{s=0} \nonumber \\ 
&-&  \frac{1}{2 \pi i} \oint_{|s|=s_0}\frac{ds}{s} \; K_{1}(s) \; \Pi_{uds}(s)  \;, \label{CAU}
\end{eqnarray}
where the integral on the right hand side, around the circle of radius
$s_{0}\simeq(1.8\;\mbox{GeV})^{2}$, is computed using PQCD in the light quark
sector. This is known up to five-loop level \cite{5L}. The contour integration can be performed using fixed order perturbation theory (FOPT) or, alternatively, contour improved perturbation theory (CIPT). There is no clear  a priori criterion to decide which is best in a given application. However, in the present case the difference between the two methods turns out to be negligible, as discussed later.
The residues are given
in terms of derivatives of the correlator at zero momentum, for which one can
use hadronic models, CHPT or lattice QCD. Hence
Eq.(\ref{AMUH1}) becomes
\begin{eqnarray}
a_\mu^{HAD}|_{uds} &=& 8 \alpha_{EM}^2 \sum_{i=u,d,s} Q_i^2 \left\{ \mbox{Res} \left[ \Pi_{uds}(s) \frac{K_{1}(s)}{s}\right]_{s=0} \nonumber \right. \\ 
&-& \left. \frac{1}{2 \pi i} \oint_{|s|=s_0}\frac{ds}{s} \; K_{1}(s) \; \Pi_{uds}(s)|_{PQCD}  
\nonumber \right. \\ 
&+& \left.
\int_{s_0}^{\infty} \, \frac{ds}{s} \; K(s) \; \frac{1}{\pi} \, \mbox{Im}\, \Pi_{uds}(s)|_{PQCD}\right\}, \label{AMUL}
\end{eqnarray}
where the last integral above involves the exact integration kernel $K(s)$ and PQCD is used for the spectral function. It is important to stress that this contribution to the anomaly only requires knowledge of a few derivatives of the vector correlator at the origin (to compute the residue). It does not require knowledge of the correlator itself in the extended energy region from threshold up to $s_{0}\simeq(1.8\;\mbox{GeV})^{2}$. The choice of this particular value for the onset of PQCD will allow for a fair comparison with determinations based entirely on data \cite{Davier}. It is also supported by experimental results from the BES Collaboration \cite{BES} which show the onset of PQCD at $s_0 \simeq 4.0\;\mbox{GeV}^2$. The stability of results against changes in this threshold value will be analyzed later.\\
 
In order to incorporate charm-quark information we add an extra contribution determined as follows. A new fit to the integration kernel $K(s)$ is performed in a region
$s_{1}\leq s\leq s_{2}$, where $s_{1}\simeq M_{J/\psi}^{2}$, and $s_{2}%
\simeq(5.0\,\mbox{GeV})^{2}$. Using this kernel and Cauchy's theorem the charm contribution is given by an expression similar to Eq.(\ref{AMUL}), except that $s_{0}$ is replaced by $s_{2}$ and $K_{1}(s)$ by the new fit function $K_{2}(s)$. The residues can now be computed directly from PQCD using the low energy expansion of the heavy-quark correlator, known up to four-loop order.
No hadronic model nor data is needed here. A similar procedure can be followed to
incorporate the contribution of the bottom quark.\newline We proceed to fit
the integration kernel, $K_{1}(s)$ in the region $s_{th}\leq s\leq s_{0}$. If
one were to choose a polynomial fit of the form $K_{1}(s)=\sum_{i=1}c_{i}%
s^{i}$ then the residues in Eq.(\ref{AMUL}) would all vanish and the anomaly
would be determined entirely from QCD \cite{SN}. There are two drawbacks to
such a fit. First, even taking many terms in the series expansion of
$K_{1}(s)$ the fit is not accurate enough. Second, the higher powers of $s$
bring in higher dimensional condensates in the operator product expansion,
thus reducing further the accuracy of this approach. An inspection of the
$s$-behavior of the kernel $K(s)$ suggests that a series expansion involving
inverse powers of $s$ should be a better option. In fact, this turns out to be the case, e.g. the fit function, Eq.(\ref{K1}), up to $s^{-3}$ becomes
\begin{eqnarray}
K_{1}(s) &=& 2.257\times 10^{-5} s + 3.482\times 10^{-3} s^{-1}\nonumber  \\ 
&-& 1.467\times 10^{-4} s^{-2} 
+ 4.722\times 10^{-6} s^{-3} \;, \label{K1N}
\end{eqnarray}
where $s$ is expressed in $\mbox{GeV}^{2}$, and the numerical coefficients
have the appropriate units to render $K_{1}(s)$ dimensionless. Figure 1 shows
the exact kernel $K(s)$ in Eq.(\ref{AMUH1}) (solid curve) together with the fit $K_{1}(s)$ as in Eq.(\ref{K1N}) (solid dots). The relative difference between the two curves lies in the range $0-1\%$ in the low energy region, where it contributes the most. A further estimate of the accuracy of the fit function, Eq.(\ref{K1N}), can be obtained by using all available experimental data on $R(s)$ in Eq.(\ref{AMUH1}) together with (a) the exact kernel Eq.(\ref{K}), and (b) the fit kernel Eq.(\ref{K1N}). We find $a_{\mu}^{HAD}|_{uds} = 641.69$ for procedure (a) and $a_{\mu}^{HAD}|_{uds} = 641.16$ for procedure (b), i.e. a difference of 0.08\%.
Using additional inverse powers of $s$ terms in the fit, while improving it slightly, it does not lead to any appreciable difference in the final result for $a_{\mu}^{HAD}$. For instance, the difference in $a_{\mu}^{HAD}$ from adding two additional inverse powers in Eq.(\ref{K1N}) is less than $0.16\%$. 

\begin{figure}[ptb]
\includegraphics[height=2.0in, width=3.0in]{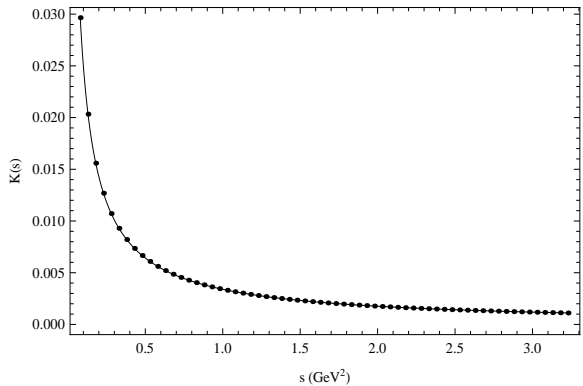}\caption{{\protect\footnotesize
{The exact kernel $K(s)$, Eq.(\ref{K}) (solid line) together with the fit in the
light-quark region, Eq.(\ref{K1N}), (solid circles). }}}%
\label{figure1}%
\end{figure}
We discuss now the incorporation of the heavy quark sector
contribution to the anomaly, starting with the charm-quark piece $a_{\mu
}^{HAD}|_{c}$. The fit to the integration kernel, which we name $K_{2}(s)$, is performed in the region $s_{1}\simeq M_{J/\psi}^{2}\leq s\leq s_{2}%
\simeq(5.0\,\mbox{GeV})^{2}$. The very simple function
\begin{equation}
K_{2}(s)=\frac{a_1}{s}\;+\;\frac{a_2}{s^2}\;,
\end{equation}
where $a_{1}=0.003712\;\mbox{GeV}^{2}$ and $a_{2}=-0.0005122\;\mbox{GeV}^{4}$,
provides an excellent fit with $K_{2}(s)$ differing from the exact kernel
$K(s)$ by less than $0.02\%$. In this case $a_{\mu}^{HAD}|_{c}$ is given by a
similar expression as $a_{\mu}^{HAD}|_{uds}$, Eq.(\ref{AMUL}), with obvious
replacements. An important difference is that now $a_{\mu}^{HAD}|_{c}$ only
involves the correlator and its imaginary part calculable in PQCD, i.e. no
hadronic model is needed for the residue. In fact, the Taylor expansion of the correlator around the origin is given by 
\begin{equation}
\Pi_{c}(s)|_{PQCD}=\frac{3}
{32\pi^{2}}\,Q_{c}^{2}\,\sum_{n\geq0}\bar{C}_{n}z^{n}\;,
\end{equation}
where $z=s/(4\overline{m}_{c}^{2})$. Here $\overline{m}_{c}\equiv\overline{m}_{c}(\mu)$ is the charm-quark mass in the $\overline{\text{MS}}$-scheme at a renormalization scale $\mu$. The coefficients $\bar{C}_{n}$ up to $n=30$ are known at three-loop level \cite{boughezal2006a}-\cite{maier2008a}. At four-loop level $\bar{C}_{0}$ and $\bar{C}_{1}$ were determined in \cite{boughezal2006a}-\cite{chetyrkin2006},
$\bar{C}_{2}$ in \cite{maier2008a} and $\bar{C}_{3}$ in \cite{maier2010}. Due
to the s-dependence of $K_{2}(s)$ no coefficients $\bar{C} 
_{4}$ and higher contribute to $\text{Res}[\Pi_{c}(s)\,p(s),s=0]$. Using as
input $\mu=3\,\mbox{GeV}$, $\alpha_{s}^{(4)}(3\,\text{GeV})=0.2145(22)$ \cite{PDG}
and $\overline{m}_{c}(3\,\text{GeV})=0.986(10)\,\text{GeV}$ \cite{mc}, we find
\begin{eqnarray}
\Pi_{c}(s)&=&0.03604+0.001833\;s+0.00012335\;s^{2}\nonumber  \\ 
&+&0.000012472\;s^{3}+\mathcal{O}(s^{4})\;,
\end{eqnarray}
where $s$ is expressed in $\mbox{GeV}^{2}$, and the numerical coefficients have the appropriate units to render $\Pi_{c}(s)$ dimensionless. The residue in the charm-quark sector is
\begin{equation}
\text{Res}\left[  \Pi_{c}(s)|_{PQCD}\frac{K_{2}(s)}{s}\right]_{s=0}
=76.1(5)\,\times10^{-7}\;,\label{RESC}%
\end{equation}
where the error is due to the uncertainty in $\alpha_{s}$ and to the
truncation of PQCD. For the bottom quark case the fit function 
\begin{equation}
K_{3}
(s)=0.003719\,\mbox{GeV}^{2}\,s^{-1}-0.0007637\,\mbox{GeV}^{4}\,s^{-2}
\end{equation}
differs from the exact kernel by less than 0.0005 \% in the range
$M_{\Upsilon}^{2}\leq s\leq(12\,\mbox{GeV})^{2}$. The residue is now
\begin{equation}
\text{Res}\left[  \Pi_{b}(s)|_{PQCD}\frac{K_{3}(s)}{s}\right]_{s=0}
=6.3\,\times10^{-7}\;,\label{RESB}%
\end{equation}
where the error is negligible. Next, in order to calculate the contour
integral around the circle of radius $s_{2}$ we make use of PQCD, i.e.
\begin{equation}
\Pi_{\text{PQCD}}(s)=\sum_{n=0}^{\infty}\left(  \frac{\alpha_{s}(\mu^{2}%
)}{\pi}\right)  ^{n}\Pi^{(n)}(s)\;,
\end{equation}
where 
\begin{equation}
\Pi^{(n)}(s)=\sum_{i=0}^{\infty
}\left(  \frac{\overline{m}^{2}}{s}\right)  ^{i}\Pi_{i}^{(n)}\;.
\end{equation}
The complete analytical result in PQCD up to $\mathcal{O}(\alpha_{s}^{2},(\overline{m}
^{2}/s)^{6})$ is given in \cite{Chetyrkin1997}, with new results up to order ${\cal{O}}(\alpha_s^2 (\overline{m}^2/s)^{30})$ obtained recently \cite{MAIER2}.
There are also exact results
for $\Pi_{0}^{(3)}$ and $\Pi_{1}^{(3)}$ from \cite{Baikov2009}, while $\Pi
_{2}^{(3)}$ is known up to a constant term \cite{Chetyrkin2000b}. This
constant term does not contribute to the contour integral due to the
s-dependence of $K_{2}(s)$. Finally, at five-loop level the
full logarithmic terms in $\Pi_{0}^{(4)}$ and $\Pi_{1}^{(4)}$ are known from
\cite{Baikov2008} and \cite{Baikov2004}, respectively. The  contour integrals in FOPT are
\begin{eqnarray}
\frac{1}{2 \pi i} \oint\frac{ds}{s} \; K_{n}(s) \; \Pi_{q}(s)|_{PQCD} \;  = \;\left\{ 
\begin{array}{lcl}
135.3 (6) \times 10^{-7} \\
\,\, \,20.3 (1) \times 10^{-7} \\
\, \,\,\,\,\,3.6 (2) \times 10^{-7}  \;,\label{OINT}
\end{array}\right.
\end{eqnarray}
for $n=1,2,3$ and $q=uds,c,b$, respectively. For $n=1$ the result in CIPT is $135.6(6) \, \times\, 10 ^{-7}$, i.e. a 0.2\% difference with FOPT. Also for $n=1$, changing the PQCD threshold in the interval
$s_0 = (1.8 - 2.0)^2\, \mbox{GeV}^2$ leads to a change of only  0.15\% in the final value of $a^{HAD}_{\mu}$. The BES Collaboration data \cite{BES} agrees well with PQCD in this region and beyond.
The results for the third
integral in Eq.(\ref{AMUL}), and their equivalent for the charm- and
bottom-quark sectors are
\begin{eqnarray}
\int_{s_j}^{\infty}  \frac{ds}{s} K(s) \, \frac{1}{\pi} \, \mbox{Im}\, \Pi_{q}(s)|_{PQCD}=\left\{ 
\begin{array}{lcl}
151.8 (1)\times 10^{-7} \\
\,\,\,20.0 (4)\times 10^{-7} \\  
\,\,\,\,\,\,3.4 (2)\times 10^{-7} \label{INT}
\end{array}\right.
\end{eqnarray}
with $j=0,2,4$ for $q=uds,c,b$, respectively. Substituting the results from
Eqs.(\ref{OINT}) and ({\ref{INT}) into Eq.(\ref{AMUL}) and the corresponding
expressions for the charm- and bottom-quark contributions, the leading order
$a_{\mu}^{HAD}$  is
\begin{eqnarray}
a_{\mu}^{HAD}&=&\frac{16}{3}\alpha_{EM}^{2}\mbox{Res}\left[  \Pi_{uds}%
(s)\frac{K_{1}(s)}{s}\right]  _{s=0} \nonumber  \\ 
&+& \, 19.4(2)\times10^{-10}\;.\label{AMUHSF}%
\end{eqnarray}
The contributions to $a_{\mu}^{HAD}$ from the charm- and bottom-quark sectors
obtained from PQCD are
\begin{equation}
a_{\mu}^{HAD}|_{c}=14.4(1)\times10^{-10}\;, \label{amuc}
\end{equation} 
and
\begin{equation}
a_{\mu}^{HAD}|_{b}=0.29(1)\times10^{-10}\;. \label{amub}
\end{equation}
Finally we discuss the calculation of the first term on the right hand side of Eq.(\ref{CAU}). Given
the parametrization in Eq.(\ref{K1N}) this term can be conveniently written
as
\begin{equation}
\mbox{Res}\left[  \Pi_{uds}(s)\frac{K_{1}(s)}{s}\right]  _{s=0}=\lim
_{s\rightarrow0}\sum_{n=1}^{3}\frac{a_{n}}{n!}\frac{d^{n}}{ds^{n}}\Pi
_{uds}(s),\label{RES}%
\end{equation}
where the $a_{n}$ are the coefficients of the $s^{-1}$, $s^{-2}$ and $s^{-3}$
terms in Eq.(\ref{K1N}), respectively.
To demonstrate the effectiveness of the method we consider three hadronic
models for the vector correlator, (single $\rho$) Vector Meson Dominance
(VMD), the Kroll-Lee-Zumino (KLZ) quantum field theory model \cite{KLZ}
-\cite{KLZ12}, and the dual resonance model realization of QCD in the large
$N_{c}$ limit ($\mbox{Dual-QCD}_{\infty}$)\cite{VEN1}-\cite{VEN3}. We reiterate that the use of hadronic models to compute the derivatives of the vector correlator at the origin is only provisional. In future, these derivatives will be provided with increased accuracy by CHPT and/or lattice QCD.
The error on VMD can be estimated to be of order $\mathcal{O}(10-20\%)$ judging from its predictions of the pion radius and form factor. The KLZ model is a renormalizable theory of
pions and a neutral $\rho$-meson which provides the necessary quantum field
theory platform for VMD, and leads to loop corrections to VMD. The loop
corrections in the KLZ model bring the pion radius and form factor into
better agreement with experiment. In VMD-type models the vector correlator
is related to the pion form factor through $\Pi_{uds}(s)\,=\,F_{\pi
}(s)/f_{\rho}^{2}$, where $f_{\rho}=4.96\pm0.02$ \cite{PDG} is the
$\gamma-\rho$ coupling and $F_{\pi}(s)$ is the pion form factor. The $\rho
$-VMD expression for the correlator is
\begin{equation}
\Pi_{uds}(s)|_{VMD}\,=\,\frac{1}{f_{\rho}^{2}}\,\frac{M_{\rho}^{2}}{(M_{\rho
}^{2}-s)}\;,
\end{equation}
which involves  the underlying standard VMD universality relation $g_{\rho\pi\pi}/f_\rho=1$. With $g_{\rho\pi\pi} = 5.92\,\pm\, 0.01$ from experiment \cite{PDG}, this relation is off by roughly 20\%.
The result for the residue in Eq.(\ref{AMUHSF}) is
\begin{equation}
\mbox{Res}\left[  \Pi_{uds}(s) K_{1}(s)/s \right]_{s=0}^{VMD}
=2.20(2)\;\times10^{-4}\;,
\end{equation}
 leading to
\begin{equation} 
a_\mu^{HAD}|_{VMD} = 644(6) \times 10^{-10}\;.
\end{equation}
For the correlator in the KLZ model we use the result from \cite{Kapusta} (see also \cite{KLZ12}) and obtain
\begin{equation}
\mbox{Res}\left[\Pi_{uds}(s) K_{1}(s)/s\right]_{s=0}^{KLZ}
= 2.22(2)\times10^{-4}\;,
\end{equation}
 and  
\begin{equation} 
a_\mu^{HAD}|_{KLZ} = 650(6) \times 10^{-10}\;.
\end{equation}
The errors for VMD and  KLZ are only those due to the uncertainty
in $f_{\rho}$ and do not include possible (systematic) model errors. The latter can be gauged from the deviation from universality $g_{\rho\pi\pi}/f_\rho=1$, off by some 20\%, as well as from the pion charge radius in VMD $<r_{\pi}^{2}>=0.394\,\mbox{fm}^{2}$, to be compared with the experimental value \cite{AMEN} $<r_{\pi}^{2}>=0.439\,\pm\,0.008\,\mbox{fm}^{2}$.\\
While QCD in the limit of an infinite number of colors leads to a hadronic
spectrum consisting of an infinite number of zero-width resonances, it does
not specify the mass spectrum nor the couplings. 
$\mbox{Dual-QCD}_{\infty}$  \cite{VEN1}-\cite{VEN3} provides this information leading to hadronic
form factors in excellent overall agreement with data in the space-like region. The vector correlator in this framework is given by
\begin{eqnarray}
\Pi_{uds}(s)|_{QCD_{\infty}} \, &=& \, \frac{1}{f_\rho^2} \, \frac{1}{\sqrt{\pi}} \frac{\Gamma(\beta - 1/2)}{\Gamma(\beta - 1)} \nonumber \\ 
&\times& \, B(\beta - 1, 1/2 - s/2 M_\rho^2) \;, \label{VEN}
\end{eqnarray}
where $\beta$ is a free parameter and $B(x,y)$
is the Euler beta-function. From the power series expansion of $B(x,y)$ it is
easy to see that Eq.(\ref{VEN}) represents an infinite number of (zero-width)
resonances corresponding to the $\rho$-meson and its radial excitations. The latter account for the deviation from the VMD result $g_{\rho\pi\pi}/f_{\rho}=1$ leading
to \cite{VEN1} $g_{\rho\pi\pi}/f_{\rho}=1.2$ in agreement with experiment{. For $\beta=2$
Eq.(\ref{VEN}) reduces to single $\rho$-VMD. The value
$\beta=2.30(3)$ results in an excellent fit to all data on the pion
form factor $F_{\pi}(s)$ in the space-like region up to $s=-10\,\mbox{GeV}^{2}$ with a chi-squared per degree of freedom $\chi_{F}
\simeq 1.5$ \cite{VEN1}. In contrast the VMD fit yields $\chi_{F}
\simeq 11$.  In addition, the {$\mbox{Dual-QCD}$}$_{\infty}$ model  gives  a value of the pion charge radius $<r_{\pi}^{2}>=0.436\,\pm
\,0.004\,\mbox{fm}^{2}$ \cite{VEN1} to be compared with the most recent experimental
value \cite{AMEN} $<r_{\pi}^{2}>=0.439\,\pm\,0.008\,\mbox{fm}^{2}$. Since the first derivative of the vector correlator dominates in Eq.(\ref{RES}), it is very important for a hadronic model to reproduce the pion radius. The result for the residue in Eq.(\ref{AMUHSF}) 
is
\begin{equation}
\mbox{Res}\left[  \Pi_{uds}(s)\frac{K_{1}(s)}{s}\right]  _{s=0}^{QCD_{\infty}}
= 2.47(3)\times10^{-4}\;, \label{lightres}
\end{equation}
 and using
Eq.(\ref{AMUHSF}) the 
hadronic $a_{\mu}^{QCD_{\infty}}$  is
\begin{equation}
a_{\mu}^{HAD}|_{QCD_{\infty}}=722\,(9)\,\times\,10^{-10}\;,\label{AMUF}
\end{equation}
where the error  is mostly due to that in $\beta$. The result Eq.(\ref{AMUF}) can be compared  with the value $a_{\mu}^{HAD}=692.3\,(4.2)\,\times\,10^{-10}$ from
\cite{Davier}-\cite{Hagiwara} using $e^{+}e^{-}$ data, or $a_{\mu}
^{HAD}=701.5(4.7)\,\times\,10^{-10}$ using $\tau$-data. However, a more recent reanalysis based on $\tau$-data \cite{Jeger2} finds
$a_{\mu}^{HAD}=690.96(4.65)\,\times\,10^{-10}$. In the $\mbox{QCD}_{\infty}$ framework the $1/N_c$ corrections arise in the time-like region from finite width resonance effects. These corrections to the form factor are of order ${\cal{O}}$$(\Gamma^2/M^2)$ near the origin. While small, they might have an impact on the residues, Eq.(\ref{lightres}). Shifting the poles in Eq.(\ref{VEN}) to the second Riemann sheet in the complex s-plane, while preserving the normalization at the origin, and the vanishing of the imaginary part of $F_\pi(s)$ at threshold \cite{VEN1},\cite{Gourdin} leads to a reasonable finite width model. The first derivative of the form factor at the origin, which is the main contribution to Eq.(\ref{lightres}),  receives no width correction. The second derivative is reduced with respect to the zero-width result by less than 2\%, and the third derivative by some 3\%. This translates into an increase in the value given in  Eq.(\ref{AMUF}) of 0.1\%.\\ 

Adding to Eq.(\ref{AMUF}) 
the QED contribution \cite{Kino} $a_\mu^{QED} = 11658471.809 \pm 0.015$, the electro-weak \cite{EW} $a_\mu^{EW} = 15.4 \pm 0.2$, the higher order hadronic \cite{Hagiwara} $a_\mu^{HAD}|_{HO} = -9.79 \pm 0.09$, and the light-by-light contribution \cite{LBL}
$a_\mu^{LbL}|_{HO} = 11.6 \pm 4.0$, all in units of $10^{-10}$, we find it intriguing that the {$\mbox{Dual-QCD}$}$_{\infty}$ \ prediction, Eq.\ref{AMUF}, leads to
\begin{equation}
a_{\mu}|_{QCD_{\infty}}=11659210.6 \pm 9.8\,\times10^{-10}\;,\label{QCDINF}
\end{equation}
to be compared with the experimental value
\begin{equation}
 a_{\mu}^{EXP}=11659208.9\pm
6.3\,\times10^{-10}\;.
\end{equation}
Equation(\ref{QCDINF}) does suggest that it might be possible to understand the muon anomaly entirely within the SM.\\
Our approach to determine $a_{\mu}^{HAD}$ appears to be optimally designed for use in CHPT, as the\ main input is the power series of the correlator around the origin. The $\mathcal{O}(p^{6})$ vector correlator was determined in \cite{bijnens2000} and in \cite{Kambor}. The derivative at zero momentum, in terms of the usual chiral constants, is
\begin{eqnarray}
\frac{d}{ds}\Pi_{uds}\bigl|_{\chi\text{pT}}(0)&=&0.0105557-4C_{93}
^{r}-0.77725L_{10}^{r} \nonumber  \\ 
&+& 1.0346L_{9}^{r}\;.
\end{eqnarray}
Two of these constants have been calculated on the lattice, $L_{9}
^{r}=3.08(23)(51)\times10^{-3}$ \cite{lattice1} and $L_{10}^{r}%
=-5.2(2)(_{-3}^{+5})\times10^{-3}$ \cite{lattice2}. The constant $C_{93}
^{r}$ has not been determined on the lattice yet. We have to
rely on a very rough estimate of this constant from \cite{bijnens2000} using
VMD, $C_{93}^{r}\approx-17\times10^{-3}\,\text{GeV}^{-2}$. With these values,
we obtain $\frac{d}{ds}\Pi_{uds}(0)\bigl|_{\chi\text{pT}}\approx0.0857$.
Given the very large uncertainty in $C_{93}^{r}$, and the fact that
the contribution of the second and  the third derivative is very small, we find
\begin{equation}
\text{Res}\Bigl [\Pi_{uds}(s)\frac{K_{1}(s)}{s}\Bigr]_{s=0} \simeq 2.8\;,
\end{equation}
leading to
\begin{equation}
a_{\mu}^{HAD} \simeq 815\times10^{-10}\;.
\end{equation}
This is a great deal larger than the value expected from experiment. The
reason for this is that the constant $C_{93}^{r}$ dominates this result. Furthermore, in \cite{bijnens2000}
it is argued that the estimates of the $\mathcal{O}(p^{6})$ constants could be larger than the physical constants. Therefore it makes sense that this result for $a_{\mu}^{HAD}$ represents an overestimate, rather than an underestimate. One can reverse this argument and give the first model independent determination of $C_{93}^{r}$. Making use of $a_{\mu}^{HAD}=692.3(4.2)$ from \cite{Davier}, we find that $C_{93}^{r}=-13.9(2)\times10^{-3}\,\text{GeV}^{-2}$.\\

In summary, we have discussed a new approach to the determination of the leading $a_{\mu}^{HAD}$ entirely from theory, i.e. without the use of experimental data on the vector correlator in an extended energy region. This can be achieved by fitting the integration kernel, Eq.(\ref{K}), in the light-quark sector with the simple function Eq.(\ref{K1N}), and subsequently invoking Cauchy's theorem in the complex s-plane. This leads to the result Eq.(\ref{AMUL}) which only requires knowledge of a few derivatives of the vector correlator at the origin. This must be contrasted with the standard approach which requires the complete correlator in the wide energy region from threshold up to $s_{0}\simeq(1.8\,\mbox{GeV})^{2}$. Such a detailed information can only be reliably and accurately  obtained from data. Currently, these derivatives can be estimated using hadronic models, examples of which have been presented here. In future, though, more accurate determinations of the derivatives from CHPT and/or lattice QCD should become available thus allowing for a model independent calculation of this contribution. In the heavy-quark sector this problem does not arise, as it is possible to calculate the anomaly entirely from PQCD, with the results given in Eqs.(\ref{amuc})-(\ref{amub}).\\

{\bf Acknowledgements}\\
This work was supported in part by NRF (South Africa) and the Alexander von Humboldt Foundation (Germany). The authors thank Dru Renner for helpful correspondence.           


\begin{thebibliography}{99}                                                  \begin{small}                  %
\bibitem {review}J. P. Miller, E. de Rafael, and B. Lee Roberts, Rep. Prog.
Phys. \textbf{70}, 795 (2007); F. Jegerlehner, and A. Nyffeler, Phys. Rep.
\textbf{477}, 1 (2009),  and references therein.
\bibitem {Davier}M. Davier \textit{et al.}, Eur. Phys. J. C \textbf{71}, 1515, (2011).
\bibitem {Hagiwara}K. Hagiwara \textit{et al.}, arXiv: 1105.3149.
\bibitem {deRK} S. J. Brodsky and E. de Rafael, Phys. Rev. \textbf{168}, 1620 (1968).
\bibitem {5L}P. A. Baikov, K. G. Chetyrkin, and J. H. K\"{u}hn, Phys. Rev.
Lett. \textbf{96}, 012003 (2006).
\bibitem{BES} J. Z. Bai {\it et al.}, BES Coll., Phys. Rev. Lett. {\bf 88}, 101802 (2002);
M. Ablikim {\it et al.}, BES Coll., Phys. Lett. B {\bf 677}, 239 (2009).
\bibitem {SN}N. Nasrallah, N. A. Papadopoulos, and K. Schilcher, Phys. Lett. B
\textbf{126}, 379 (1983).
\bibitem {boughezal2006a}R. Boughezal, M. Czakon, and T. Schutzmeier, Phys.
Rev. D \textbf{74}, 074006 (2006).
\bibitem {maier2008a}A. Maier, P. Maier\"{o}fer, and P. Marquard, Nucl. Phys.
B \textbf{797}, 218 (2008); Phys. Lett. B \textbf{669}, 88 (2008).
\bibitem {chetyrkin2006}K. G. Chetyrkin, J. H. K\"{u}hn, and C. Sturm, Eur.
Phys. J. C \textbf{48}, 107 (2006).
\bibitem {maier2010}A. Maier \textit{et al.}, Nucl. Phys. B \textbf{824}, 1 (2010).
\bibitem {PDG}K. Nakamura \textit{et al.}, Particle Data Group, J. Phys. G
\textbf{37}, 075021 (2010).
\bibitem {mc}K. G. Chetyrkin \textit{et al.}, Phys. Rev. D \textbf{80}, 074010
(2009); S. Bodenstein \textit{et al.}, Phys. Rev. D \textbf{82}, 114013
(2010); \textit{ibid.} \textbf{83}, 074014 (2011).
\bibitem {Chetyrkin1997}K. G. Chetyrkin \textit{et al.}, Nucl. Phys. B
\textbf{503}, 339 (1997).
\bibitem{MAIER2} A. Maier, and P. Marquard, arXiv:1110.558.
\bibitem {Baikov2009}P. A. Baikov, K. G. Chetyrkin, and J. H. K\"{u}hn, Nucl.
Phys. B (Proc. Suppl.) \textbf{189}, 49 (2009).
\bibitem {Chetyrkin2000b}K. G. Chetyrkin, R. Harlander, and J. H. K\"{u}hn,
Nucl. Phys. B \textbf{586}, 56 (2000).
\bibitem {Baikov2008}P. A. Baikov, K. G. Chetyrkin, and J. H. K\"{u}hn, Phys.
Rev. Lett. \textbf{101}, 012002 (2008).
\bibitem {Baikov2004}P. A. Baikov, K. G. Chetyrkin, and J. H. K\"{u}hn, Nucl.
Phys. B (Proc. Suppl.) \textbf{135}, 243 (2004).
\bibitem {KLZ}N. M. Kroll, T. D. Lee, and B. Zumino, Phys. Rev. \textbf{157},
1376 (1967).
\bibitem {KLZ12}C. A. Dominguez \textit{et al.}, Phys. Rev. D \textbf{76},
095002 (2007); C. A. Dominguez, M. Loewe, and B, Willers, Phys. Rev. D
\textbf{78}, 057901 (2008).
\bibitem {VEN1}C. A. Dominguez, Phys. Lett. B \textbf{512}, 331 (2001); C. A.
Dominguez, and T. Thapedi, J. High Ener. Phys. \textbf{0410}, 003 (2004); C.
A. Dominguez, and R. R\"{o}ntsch, \textit{ibid.} \textbf{0710}, 085 (2007).
\bibitem{VEN2} E. Ruiz Arriola and W. Broniowski, Phys. Rev. D {\bf 78}, 034031 (2008);
{\it ibid.} D {\bf 81}, 094021 (2010); W. Broniowski and E. Ruiz Arriola, PoS LC 2010:062 (2010).
\bibitem{VEN3} C. Bruch, A. Khodjamirian, and J.H. K\"{u}hn, Eur. Phys. J. C {\it 39}, 41 (2005).
\bibitem {Kapusta}C. Gale and J. I. Kapusta, Nucl. Phys. B {\bf 357}, 65 (1991).
\bibitem {AMEN}S. R. Amendolia \textit{et al.}, Nucl. Phys. B \textbf{277},
168 (1986).
\bibitem{Jeger2} F. Jegerlehner and R. Szafron, Eur. Phys. J. C {\bf 71}, 1632 (2011).
\bibitem{Gourdin} M. Gourdin, Phys. Rep. C {\bf 11}, 29 (1974).
\bibitem {Kino}T. Kinoshita and M. Nio, Phys. Rev. D \textbf{73}, 013003 (2006).
\bibitem {EW}A. Czarnecki, W. J. Marciano, and A. Vainshtein, Phys. Rev. D
\textbf{67}, 073006 (2003); M. Knecht \textit{et al.}, J. High Ener. Phys.
{\b0211}, 003 (2002).
\bibitem {LBL}A. Nyffeler, Phys. Rev. D \textbf{79}, 073012(2009).
\bibitem {bijnens2000} J. Bijnens, P. Talavera, Nuclear Physics B
568 (2000) 319.
\bibitem {Kambor}E. Golowich and J. Kambor, Nucl. Phys. B447 (1995) 373
\bibitem {lattice1} P. A. Boyle, et. al., JHEP 07 (2008) 112.
\bibitem {lattice2} E. Shintani, et. al., Phys. Rev. Lett.
\textbf{101} (2008) 242001.
\end{small} 
\end{thebibliography}
\end{document}